\documentclass[aps,12pt,pra,showpacs,superscriptaddress,preprint]{revtex4}
\usepackage{amsmath}
\usepackage{graphicx}
\usepackage{subfig}
\usepackage{array}
\usepackage{amsfonts}

\catcode`ð=\active
\defð{\u{g}}
\catcode`Ð=\active
\defÐ{\u{G}}
\catcode`Ý=\active
\defÝ{\. I}
\catcode`ö=\active
\defö{\"{o}}
\catcode`Ö=\active
\defÖ{\"O}
\catcode`ü=\active
\defü{\"{u}}
\catcode`Ü=\active
\defÜ{\"{U}}
\catcode`Þ=\active
\defÞ{\c{S}}
\catcode`þ=\active
\defþ{\c{s}}
\catcode`ý=\active
\defý{{\i}}
\catcode`ç=\active
\defç{\d{c}}
\catcode`Ç=\active
\defÇ{\d{C}}

\begin{document}

\title{Exact Solution of a Spin-$\frac{1}{2}$ Particle for a Linear Potential}
\author{\small Altuð Arda}
\email[E-mail: ]{arda@hacettepe.edu.tr}\affiliation{Department of
Physics Education, Hacettepe University, 06800, Ankara,Turkey}

\begin{abstract}
The problem of a spin-$\frac{1}{2}$ particle moving in a linear
potential field in two-dimensions is searched to obtain for
nonzero energy eigenvalues and the corresponding normalized
eigenfunctions. The zero-mode ($E=0$) eigenfunctions are also
studied and it is seen that they are normalizable. The variation
of the non-zero eigenfunctions and also eigenvalues
are according to the position and potential parameter $\gamma$, respectively, given in the text.
\end{abstract}

\maketitle

\newpage

\section{Introduction}
The potential having a linear form of position and a
"Coulomb-like" potential play an important role in various
branches of physics. The gluon condensation in high energy physics
can be studied with an effective linear potential [1]. The
different properties of low-lying baryons have been searched by
using a linear potential and the corrections to the mass spectrum
have also been calculated one-gluon exchange [2]. It has been
suggested that the hadron spectrum can be studied by assuming that
the quarks are bounded by a long-range potential plus a
short-range potential (a Coulomb-like) arising in gluon-photon
exchange diagrams [3]. It is obtained that baryon resonances can
be studied by using a gluon-perturbed linear potential. In
addition, it has been analytically showed that the static QCD
potential can be designed as a sum of a "Coulomb-like" potential
and a linear potential by using an approximation based on the
renormalon picture [4]. The linear potential arising from an
effective vector exchange can be used to calculate fine-structure
corrections $\psi/J$ particles and also compute the
electromagnetic decay rates between low-lying $s$ and $p$ states
[5].

Analytical solutions of Dirac equation for a "Coulomb-like"
potential and/or a linear potential or their linear combination
are also a subject within the relativistic quantum mechanics
[6-8].

In the present work the exact solutions of two-dimensional Dirac
equation for a particle moving in a linear potential are obtained
by turning it into a second order cylinder differential equation
[9]. The nonzero eigenfunctions are written in terms of the
functions $\,_{1}F_{1}(a; b; z)$ and then they are given in terms
of the Hermite polynomials. The nonzero eigenenergies are obtained
by using some restrictions on the eigenfunctions. The
normalization constants are also computed and the variation of the
first three eigenfunctions according to the position are given in
plots. The zero-mode eigenenergies ($E=0$) and the corresponding
eigenfunctions are computed from the Dirac equation.

\section{Analytical Solutions}

The (1+1)-dimensional time-independent Dirac equation for a
spin-$\frac{1}{2}$ particle with rest mass $m$ moving in a
time-independent scalar potential is given
\begin{eqnarray}
\left[c\vec{\alpha}.\vec{p}+\beta\left(mc^2+V(x)\right)\right]\Psi(x)=E\Psi(x)\,,
\end{eqnarray}
where $\vec{p}$ is the momentum operator and $\vec{\alpha}$ and
$\beta$ are Hermitian square matrices, respectively. $E$ is the
energy of particle and $c$ is the velocity of light. Choosing
$\vec{\alpha}$ and $\beta$ as $2 \times 2$ Pauli spin matrices as
$\vec{\alpha}=\sigma_{2}$ and $\beta=\sigma_{2}$ and writing the
spinor in terms of the upper and lower components as
\begin{eqnarray}
\Psi(x)=\begin{pmatrix}
 \psi(x) \\
 \phi(x)
\end{pmatrix}\,,
\end{eqnarray}
the Dirac equation reads as
\begin{subequations}
\begin{align}
[mc^2+V(x)]\phi(x)-\hbar c\frac{\phi(x)}{dx}&=E\psi(x)\,,\\
[mc^2+V(x)]\psi(x)+\hbar c\frac{\psi(x)}{dx}&=E\phi(x)\,,
\end{align}
\end{subequations}

Last two equations can be written as two second order differential
equations
\begin{subequations}
\begin{align}
\frac{d^2\psi(x)}{dx^2}+\mathop{U_{u}}_{l}(x)\psi(x)&=\mathop{\epsilon_{u}}_{l}\psi(x)\,,\\
\frac{d^2\phi(x)}{dx^2}+\mathop{U_{u}}_{l}(x)\phi(x)&=\mathop{\epsilon_{u}}_{l}\phi(x)\,,
\end{align}
\end{subequations}
where $u$ stands for "upper" and $l$ for "lower", respectively and
\begin{subequations}
\begin{align}
\mathop{U_{u}}_{l}(x)&=-\frac{1}{\hbar^2c^2}V^2(x)-\frac{2m}{\hbar^2}V(x)\pm\frac{1}{\hbar c}\frac{dV(x)}{dx}\,,\\
\mathop{\epsilon_{u}}_{l}&=\frac{1}{\hbar^2c^2}(E^2-m^2c^4)\,,
\end{align}
\end{subequations}

It should be noted that Eqs. (4a) and (4b) are not independent
because of the eigenvalue $\mathop{\epsilon_{u}}_{l}$. So, we have
to look for bound state solutions for $\mathop{U_{u}}_{l}(x)$
having a common energy eigenvalue.

Now let us consider a scalar potential with the following form
\begin{eqnarray}
V(x)=-V_{0}-\gamma x\,,
\end{eqnarray}
where $\gamma$ is a real parameter. With the help of this equation
and using Eq. (5a), we obtain the following from Eq. (4a)
\begin{eqnarray}
\left\{\frac{d^2}{dx^2}-\left(\alpha_{1}x^2+\alpha_{2}x+\alpha_{3}\right)\right\}\psi(x)=0\,,
\end{eqnarray}
where
\begin{eqnarray}
\alpha_{1}=\frac{\gamma^2}{\hbar^2c^2}\,\,;\alpha_{2}=\frac{2\gamma}{\hbar^2}\left(\frac{V_{0}}{c^2}-m\right)\,\,;
\alpha_{3}=\frac{V_{0}}{\hbar^2}\left(\frac{V_{0}}{c^2}-2m\right)+\frac{\gamma}{\hbar
c}+\frac{1}{\hbar^2c^2}(E^2-m^2c^4)\,,
\end{eqnarray}
Eq. (8) is written in terms of a new variable
$y=\left(4\alpha_{1}\right)^{1/4}x$ as
\begin{eqnarray}
\left\{\frac{d^2}{dy^2}-\left(\frac{1}{4}y^2+\frac{\alpha_{2}}{\left(4\alpha_{1}\right)^{3/4}}y
+\frac{\alpha_{3}}{\left(4\alpha_{1}\right)^{1/4}}\right)\right\}\psi(y)=0\,,
\end{eqnarray}
In order to obtain a cylinder differential equation [9] we use a
new variable
$z=y+\frac{2\alpha_{2}}{\left(4\alpha_{1}\right)^{3/4}}$ in last
equation and then we obtain
\begin{eqnarray}
\frac{d^2\psi(z)}{dz^2}-\left(\frac{1}{4}z^2+A\right)\psi(z)=0\,,
\end{eqnarray}
where
$A=\frac{\alpha_{3}}{\left(4\alpha_{1}\right)^{1/2}}-\frac{\alpha_{2}}{\left(4\alpha_{1}\right)^{3/2}}$.
For the convenience, we want to write this equation as following
\begin{eqnarray}
\left(\frac{d^2}{dz^2}-\frac{1}{4}z^2\right)\psi(z)=\frac{1}{2}(B+1)\psi(z)\,,
\end{eqnarray}
where $B=\frac{1}{\hbar c\gamma}(E^2-m^2c^4)$.

The general solutions of Eq. (10) are written in terms of a
confluent hypergeometric function [9]
\begin{subequations}
\begin{align}
\psi(z)&\sim e^{-z^2/4}\,_{1}F_{1}\left(\frac{A}{2}+\frac{1}{4}; \frac{1}{2}; \frac{1}{2}z^2\right)\,,\\
\psi(z)&\sim ze^{-z^2/4}\,_{1}F_{1}\left(\frac{A}{2}+\frac{1}{4};
\frac{1}{2}; \frac{1}{2}z^2\right)\,,
\end{align}
\end{subequations}
corresponding to "odd" and "even" solutions, respectively. In
order to get a finite (physical) solutions we have to write
following equality in Eq. (12a)
\begin{eqnarray}
\frac{A}{2}+\frac{1}{4}=-n\,\,\,\,\,(n \in \mathbb{N})\,,
\end{eqnarray}
which gives a quantization condition for the energy eigenvalues.
Using the last restriction, we write the eigenfunctions for $x>0$
\begin{eqnarray}
\psi(z)_{x>0}=Ne^{-z^2/4}\,_{1}F_{1}(-n; \frac{1}{2};
\frac{1}{2}z^2)\,,
\end{eqnarray}
where $N$ is obtained from the normalization. The requirement in
Eq. (13) implies into energy eigenvalues
\begin{eqnarray}
E=\pm\sqrt{2n'\hbar c\gamma+2m^2c^4-\hbar c\gamma\,}\,,
\end{eqnarray}
where $n'=2n+\frac{1}{2}$. It is seen that the energy levels are
symmetric about $E=0$ and proportional to the potential parameter
$\gamma$.

We could write the eigenfunctions in Eq. (14) in a more suitable
form by using the following identity between confluent
hypergeometric function and the Laguerre polynomials
$L_{n}^{m}(x)$ [9]
\begin{eqnarray}
\,_{1}F_{1}(-n; m+1; x)=\frac{m!n!}{(m+n)!}L_{n}^{m}(x)\,,
\end{eqnarray}
and the identity between the Laguerre polynomials and the Hermite
ones $H_{n}(x)$
\begin{eqnarray}
L^{-1/2}_{n}(x)=\frac{(-1)^{n}}{2^{2n}n!}H_{2n}(\sqrt{x\,}\,)\,,
\end{eqnarray}
With the help of last two equations, we write the eigenfunctions
as
\begin{eqnarray}
\psi(z)_{x>0}=N\frac{(-1)^{n}(n-1)!}{(2n-1)!}\,e^{-z^2/4}H_{2n}(\sqrt{\frac{1}{2}\,}z)\,,
\end{eqnarray}
where the normalization constant is obtained from
$\int_{-\infty}^{+\infty}(|\psi|^2+|\phi|^2)dx=1$.

Following the same steps we collect the differential equations
satisfying the eigenfunctions $\psi(x)$ and $\phi(x)$ as
\begin{subequations}
\begin{align}
\left(\frac{d^2}{dz^2}-\frac{1}{4}z^2\right)\psi(z)_{x>0}&=\frac{1}{2}(B+1)\psi(z)_{x>0}\,,\\
\left(\frac{d^2}{dz^2}-\frac{1}{4}z^2\right)\phi(z)_{x>0}&=\frac{1}{2}(B-1)\phi(z)_{x>0}\,,
\end{align}
\end{subequations}
and
\begin{subequations}
\begin{align}
\left(\frac{d^2}{dz^2}-\frac{1}{4}z^2\right)\psi(z)_{x<0}&=\frac{1}{2}(B-1)\psi(z)_{x<0}\,,\\
\left(\frac{d^2}{dz^2}-\frac{1}{4}z^2\right)\phi(z)_{x<0}&=\frac{1}{2}(B+1)\phi(z)_{x<0}\,,
\end{align}
\end{subequations}
These equations shows that the eigenfunctions could be constructed
from Hermite polynomials of order $2n$ and $2n+1$. On the other
hand, the continuity condition at $x=0$ gives the requirement that
$H_{2n+1} \sim const.\,H_{2n}$ where the "const." could be
determined from the continuity condition [7].

Here we should to say that the energy eigenvalues for $x<0$ is
little different from the result given in Eq. (15). The quantized
energy values for $x<0$ is given as
\begin{eqnarray}
E=\pm\sqrt{2n'\hbar c\gamma+2m^2c^4+\hbar c\gamma\,}\,,
\end{eqnarray}
which shows that the energy levels are also symmetric about $E=0$.
The difference between two results obtained from Eqs. (15) and
(21) is plotted in Fig. (1). It is seen that the dependence of the
eigenvalues on the parameter $\gamma$ is linear while the
contribution of $\gamma$ to the energy values is lower in the case
given in Eq. (21) than the ones given in Eq. (15).

In order to get the normalization constant in Eq. (14) we use the
following identity for the Hermite polynomials [9]
\begin{eqnarray}
\int_{-\infty}^{+\infty}H_{n}(x)H_{m}(x)e^{-x^2}dx=2^{n}(n!)\sqrt{\pi\,}\delta_{mn}\,,
\end{eqnarray}
and we obtain as
\begin{eqnarray}
N=\frac{1}{(n-1)!(-1)^{n}}\,\sqrt{\frac{(2n-1)!}{n2^{2n-1}\sqrt{\pi\,}}\,}\,,
\end{eqnarray}

The dependence of the normalized eigenfunctions on the coordinate
$z$ with the help of the last equation could be seen in Fig. (2).
All eigenfunctions have the same behavior near the origin and for
$z \rightarrow +\infty$ and also finite values. They oscillate
between the range where they are well defined.

Finally let us search the zero-mode solutions which can be
obtained from Eq. (3). In this case we obtain first-order
differential equations and the solutions could be summarized as
\begin{subequations}
\begin{align}
\psi(x)_{x>0}&\sim e^{h^{x>0}(x)}\,,\\
\phi(x)_{x>0}&\sim e^{-h^{x>0}(x)}\,,\\
\psi(x)_{x<0}&\sim e^{h^{x<0}(x)}\,,\\
\phi(x)_{x<0}&\sim e^{-h^{x<0}(x)}\,,
\end{align}
\end{subequations}
where
\begin{subequations}
\begin{align}
h^{x>0}(x)&=\frac{\gamma^2}{2\hbar c}x^2+\frac{1}{\hbar c}(V_{0}-mc^2)x\,,\\
h^{x<0}(x)&=-\frac{\gamma^2}{2\hbar c}x^2+\frac{1}{\hbar
c}(V_{0}-mc^2)x\,,
\end{align}
\end{subequations}
These solutions can be normalized by using the following identity
[9]
\begin{eqnarray}
\int_{-\infty}^{+\infty}e^{-(ax^2+bx+c)}dx=\sqrt{\frac{\pi}{a}\,}exp\left[\frac{b^2-4ac}{4a}\right]\,,\,\,(a>0)
\end{eqnarray}
and then we find the normalization constant $N'$ as
\begin{eqnarray}
N'=\left[\sqrt{\frac{\pi}{\sqrt{-\frac{\gamma^2}{\hbar c
}\,}}\,}exp\left[-\frac{1}{\hbar
c\gamma}(mc^2-V_{0})^2\right]\right]^{-1/2}\,,
\end{eqnarray}

We have completely analyzed the problem of a spin-$\frac{1}{2}$
particle subject to a linear potential field. We have computed the
energy eigenvalues and the corresponding normalized eigenfunctions
by converting the Dirac equation to the second-order cylinder
differential equation. We have written the eigenfunctions by using
the Laguerre and the Hermite polynomials [7]. We have seen that
the upper and lower components of the Dirac spinor are the same
expect a constant which could be determined from the continuity
condition at $x=0$. We have computed the probability density
$|\Psi|^2$ and $|\psi|^2$ and numerically showed their variation
with respect to $x$ in figure. We have also searched the zero-mode
energy eigenvalues and the corresponding normalized wave
functions.

\newpage

\begin{figure}
\centering \subfloat{
\includegraphics[height=2.2in,width=2.8in]{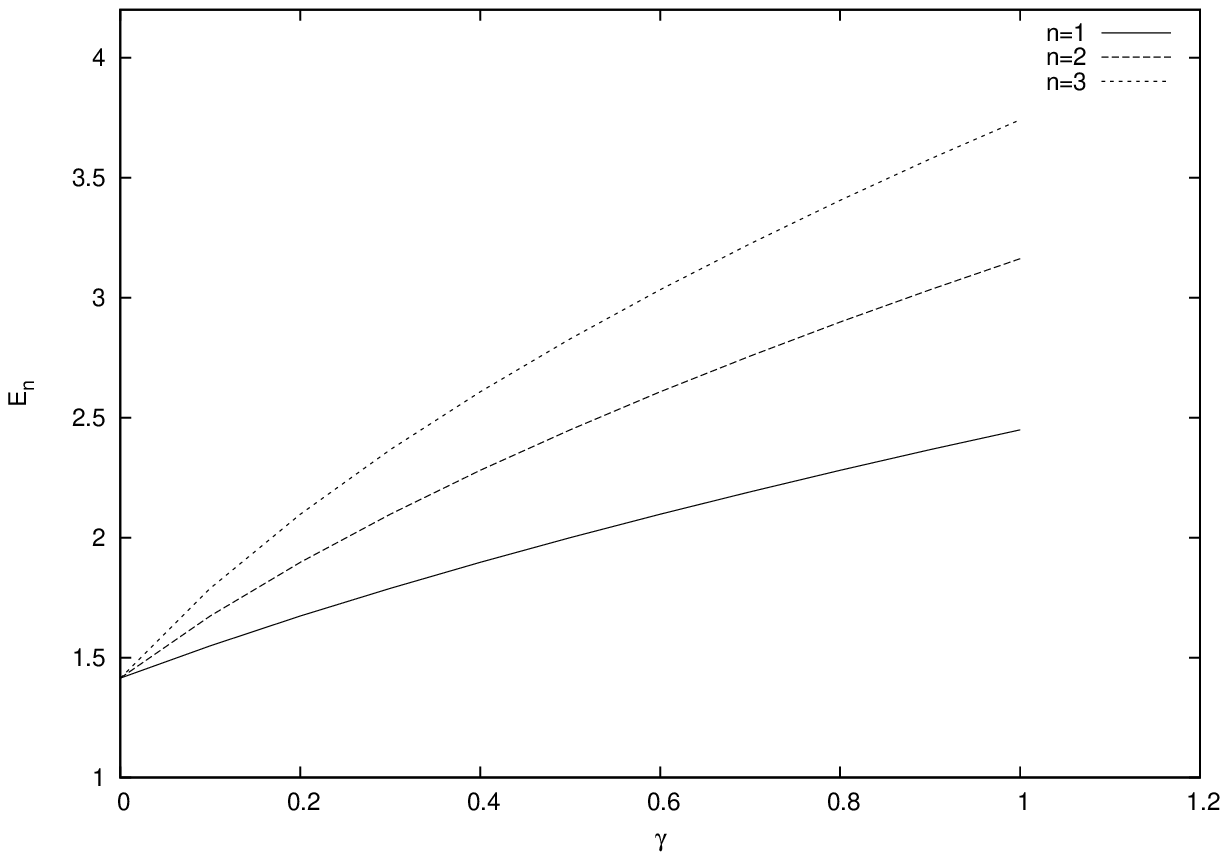}}
\hspace{0.1\linewidth} \subfloat{
\includegraphics[height=2.2in,width=2.8in]{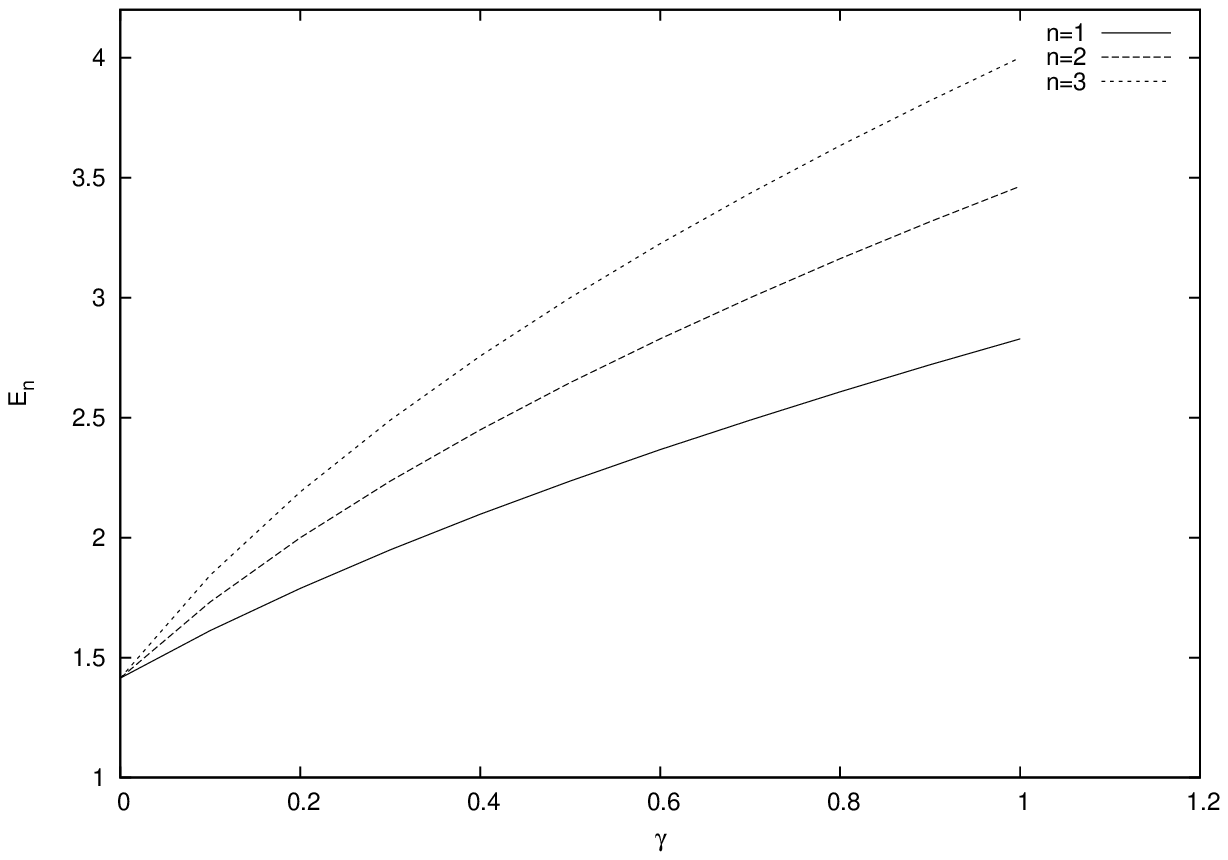}}
\caption{Variation of the first three positive energy eigenvalues
with the potential parameter $\gamma$ obtained from Eq. (15) (left
panel) and from Eq. (21) (right panel) ($m=c=\hbar=1$).}
\end{figure}

\newpage

\begin{figure}
\centering \subfloat{
\includegraphics[height=2.2in,width=2.8in]{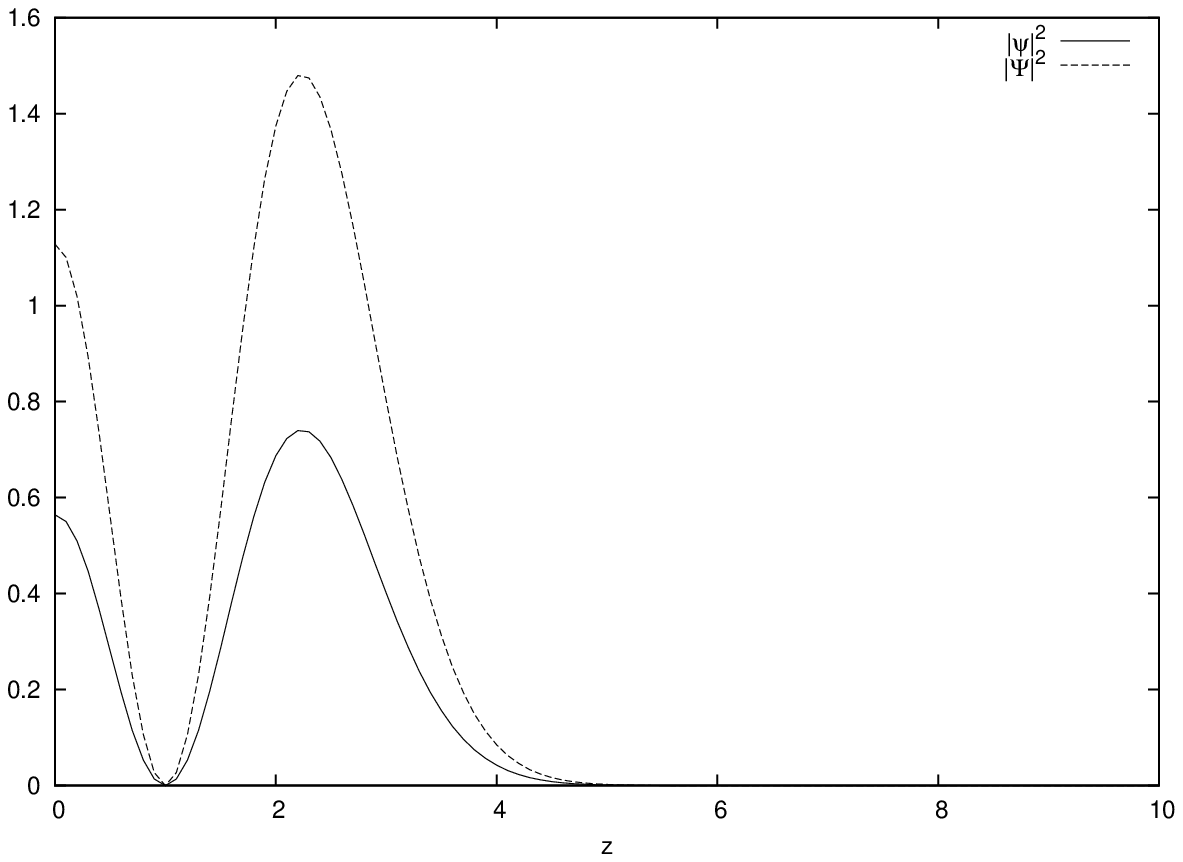}}
\hspace{0.1\linewidth} \subfloat{
\includegraphics[height=2.2in,width=2.8in]{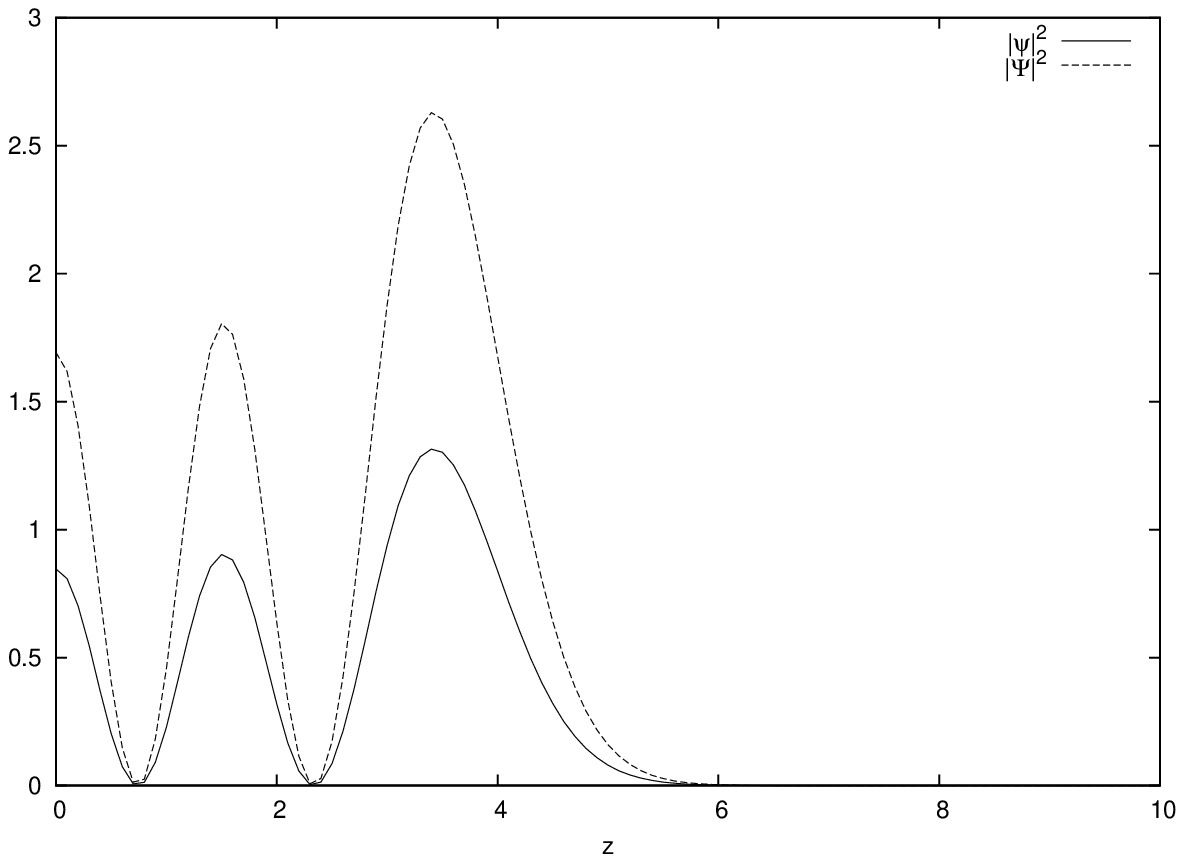}}
\vspace{0.1\linewidth} \subfloat{
\includegraphics[height=2.2in,width=2.8in]{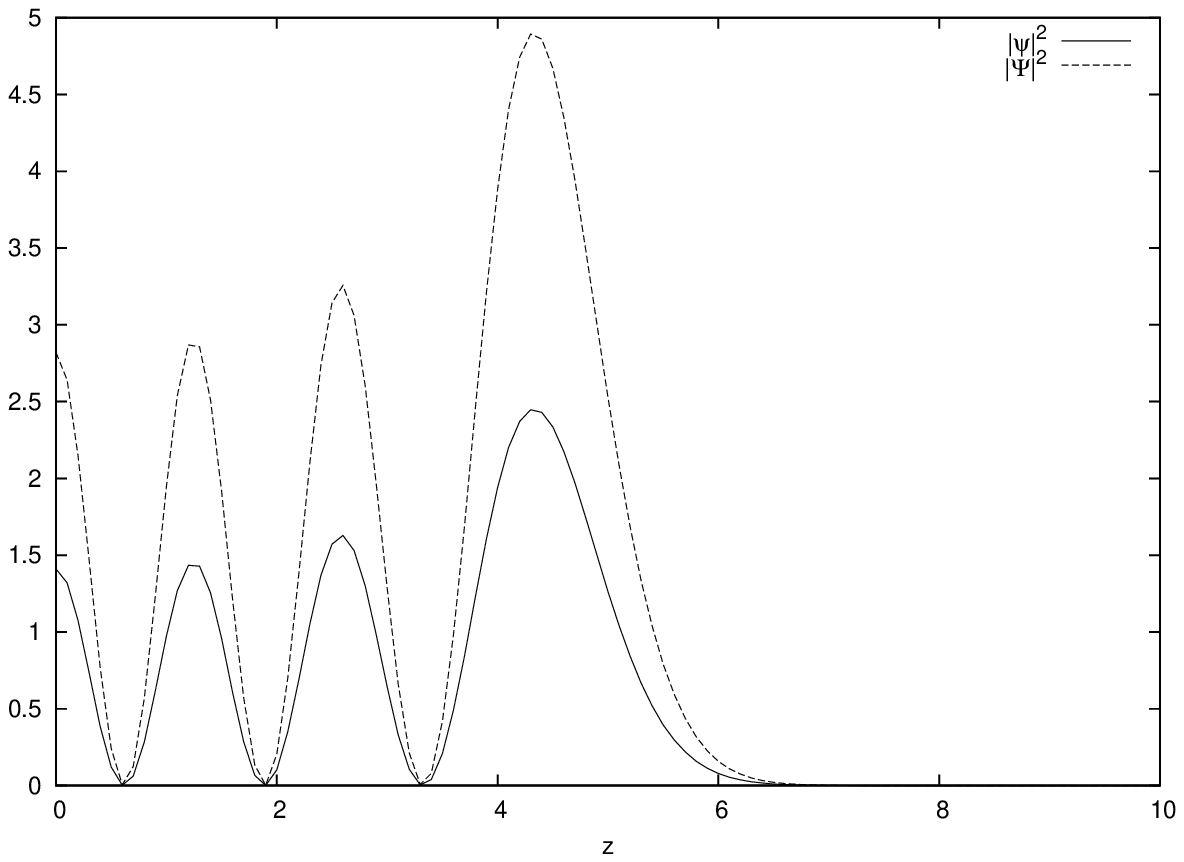}}
\caption{Variation of eigenfunctions according to $z$ for $n=1$
(left upper panel), $n=2$ (right upper panel) and $n=3$ (lower
panel) ($m=c=\hbar=1$).}
\end{figure}

\end{document}